# Twisted domains in ferroelectric nematic liquid crystals

Vladimíra Novotná, Lubor Lejček and Natalia Podoliak,

*Institute of Physics of the Czech Academy of Sciences, Na Slovance 1999/2, 182 00 Prague 8, Czech Republic*

Corresponding author*: Vladimíra Novotná, e-mail: novotna@fzu.cz

Key words: liquid crystals, ferroelectric nematics, defects, twisted domains

**Abstract**

Ferroelectric nematics present properties promising for applications, which strongly depend on the confinement conditions and the presence of defects. In this work, we have observed and analysed textures in the cells with homogeneous geometry for a ferroelectric nematogen. The studied compound reveals the ferroelectric nematic phase, $N_F$, directly on cooling from the isotropic phase. The $N_F$ phase is stable and persists down to the room temperatures. Textures in the homogeneous cell with antiparallel alignment prefer to be arranged in twisted domains with borderlines oriented preferentially in the perpendicular direction with respect to the surface alignment (rubbing direction). We have described these domains and developed a model of $2\pi$ disclinations, which are separating the areas with an opposite twist sense. We discussed the character of disclination charge and explained why these disclination lines prefer to be oriented perpendicularly to the rubbing direction.

## 1      Introduction

Since its discovery, the ferroelectric nematic phase, $N_F$, has started to be intensively studied because of its immense potential due to very large polarisation and big dielectric susceptibility values. Even though this phase was predicted more than hundred years ago [1], the first experimental examples appeared in 2017 when the ferroelectric properties of the $N_F$ phase were proven for two compounds [2-5] designated DIO and RM734. Since this revolutionary achievement, the scientists from the field of liquid crystals (LCs) try to investigate the $N_F$ phase from various aspects [6-10]. The physical properties of both compounds were intensively studied and high values of polarisation and permittivity were reported.[11] The polarisation up to 5 μC/cm$^2$ is for an order higher value than the values previously reported for other LC phases [12-15]. Additionally, large permittivity values evidence of strong sensitivity to an external field.

Defects are present in any LCs in confined geometries depending on the boundary conditions, external stimuli, elastic constants, or surface conditions. Two main cell geometries can be introduced with respect to the anchoring conditions at the surfaces: planar (HG) or homeotropic (HT) alignment, when molecules are parallel or perpendicular to the cell surface, respectively. In the case of polar molecules in a homogeneous geometry (HG), parallel or antiparallel alignment at opposite surfaces can be arranged by rubbing the surfactant layer. Previously, a careful analysis was devoted to various textures observed under a polarising microscope in different geometries for both studied compounds DIO and RM734, see [2-10]. The existence of ferroelectric domains with a different sense of dipolar twist in neighbouring domains has been reported [8,9,13]. Nevertheless, there have been only several works theoretically describing the ferroelectricity in the nematic phase [16-19]. Ferroelectric nematics represent polar liquids and the topological aspect is remarkable. Several recent publications have been devoted to defects and their description [20] with a conclusion that there is a variety of topological defects, which are present in any type of the cells [21,22].

The interaction of the molecules with the surfaces in the $N_F$ phase is a key aspect for further applications of these new compounds. At this moment, there is only a limited number of papers dealing with defects as topological singularities of the molecular arrangement in the $N_F$ phase and more data as well as a theoretical explanation is necessary. Recently, we presented new molecular structures [23] revealing the $N_F$ phase in a broad temperature range observed directly on cooling from the isotropic phase. The advantage of our compounds lies in their thermal stability and strong ferroelectricity persisting to the room temperatures. In this contribution, we have investigated textural features of the $N_F$ phase in the HG geometry with the antiparallel rubbing. We observed at least two different types of domains and herein we describe their character. Additionally, the theoretical model is presented targeted at the behaviour of the defects separating such twisted domains.

## 2      Results

**Experimental observations of textures in $N_F$ sample**

For textural observations we utilised commercial cells with homogeneous antiparallel alignment (HG cells), which are composed of glass plates covered with transparent ITO electrodes and surfactant layers. The surfactant layers ensure an antiparallel rubbing resulting in the opposite director orientations at the cell surfaces. The molecular director is connected with the polarisation, i.e. $\vec{n} \sim \vec{P}$, and the director orientations on the upper and lower surfaces differ by $\pi$ in a cell with antiparallel rubbing [8,13]. Therefore, the transition to the $N_F$ phase is accompanied with the appearance of the director twist rotation within the sample with the possibility of two opposite orientations (right-handed and left-handed). The probability of these two orientations is equal, thus, the structure of the neighbouring domains of opposite twist (twisted domains) is formed. Both twisted domains have the same surface orientations on the upper and the lower sample surfaces. We assume that the boundary between the twisted domains with the opposite rotation does not have a character of a wall starting and finishing on the sample surfaces. Such a domain wall would not be stable and probably would collapse into $2\pi$-twist disclination separating domains with an opposite twist sense [13]. Due to these reasons, we consider that the domain boundaries are 2π-twist disclination lines within the sample bulk.

For the studied compounds [23] we have observed large homogeneous areas with twisted domains for all HG cells with antiparallel rubbing under an optical polarising microscope with crossed polariser and analyser. Indeed, for a thin 1.6 μm HG cell, an individual domain size can reach up to 1mm². In this paper, we present the observations of the twist domain boundaries in the $N_F$ phase on a 5 μm cell with antiparallel rubbing (Fig. 1). Let us point out that the elongated parts of domains are generally oriented perpendicularly to the rubbing direction. The twisted domains can be distinguished when rotating the analyser from the crossed position to an angle of about 10 degrees clockwise or anticlockwise (Fig. 2). The boundaries between domains, which correspond to 2π-twist disclinations, are decorated with point-like objects (Fig. 1b and 1c in an enlarged view).

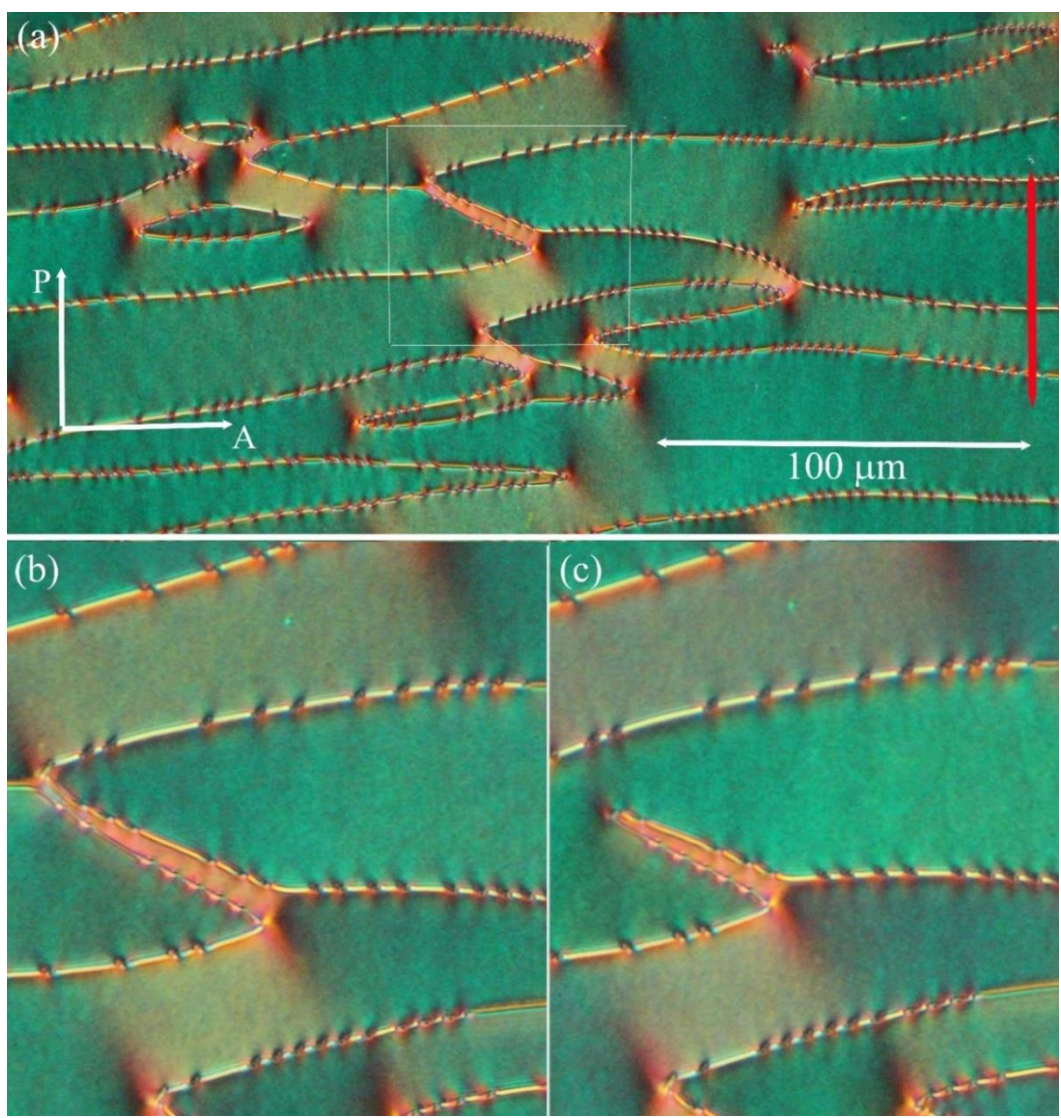

**Fig. 1.** Planar textures in a 5 μm HG antiparallel cell show twist domains, which are preferably elongated perpendicular to the rubbing direction (red arrow). In (a) there is a larger area with disclination segments, in (b) a segment between two domains with the same twist starts to annihilate and (c) shows the continuous annihilation and the boundary between the domains is ceasing to exist.

The disclination segments between the neighbouring twist domains can start to annihilate, as it is demonstrated in Fig. 2. Within a temperature stabilisation up to 20 seconds, one part of the disclination line showed a tendency to shorten the domain circumference, as it is demonstrated in Fig. 1b and Fig. 1c for comparison. This means that the neighbouring segments have the opposite topological charges. Such an effect is demonstrated in detail in Fig.2, in which the situation is presented for the analyser rotated at an angle of about 10 degrees from the crossed orientation with respect to the polariser. Under this condition, both types of domains are clearly distinguishable. In a central part of the figure, we were able to observe an elongated domain continuously disappearing. Boundaries between twist domains, which correspond to 2π-twist disclinations, are decorated by point-like defects. The point-like objects also disappear when the disclination line segments annihilate. It means that these objects are connected with the presence of disclinations and they represent the deformations of the director field.

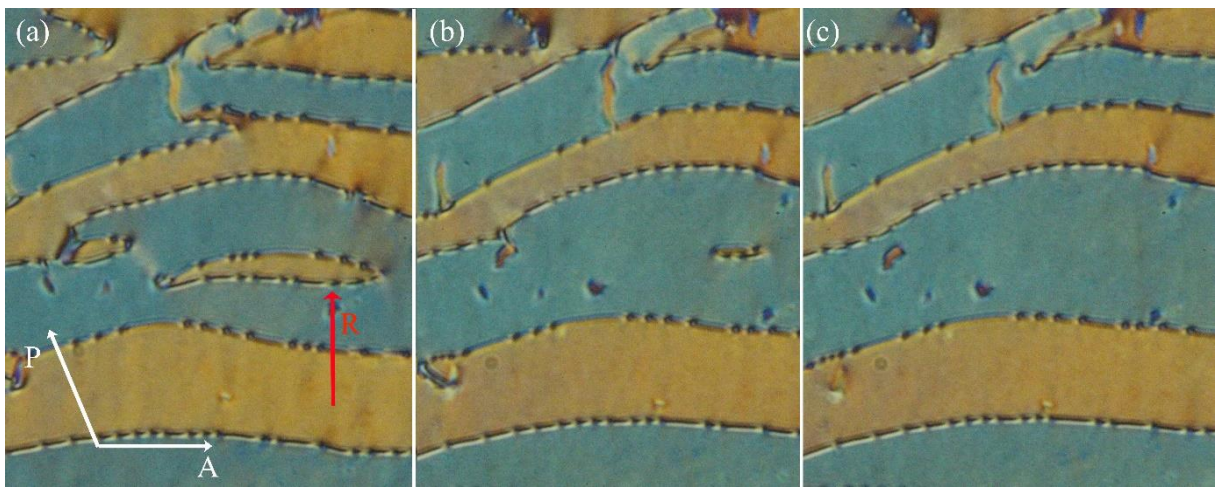

**Fig. 2.** Sequence of photos documenting the time-development of a twisted domain in a 5 μm HG antiparallel cell. The analyser is rotated at an angle of about 10 degrees from a crossed position. A red arrow is oriented along the rubbing direction and marks the observed domain. The circumference of the domain tends to shorten during the temperature stabilisation: (a) at a starting time t=0, (b) after 10s, the photo shows a small remaining part of the disclination just before its total annihilation, (c) the situation after 20 seconds. Note that point-like defects, which were present on the disclination line, also annihilated.

For the utilised HG cells, an electric field can be applied perpendicularly to the glass plates. In such alignment of the sample, the polarisation vector is perpendicular to the applied field and a clear assessment of the field effects is not easy. We observed rather complex modifications of the textures when the electric field (of about 20 V) was applied to the HG antiparallel cell (5 μm). Under the electric field, we observed a change of the twisted domains shape and colour. Nevertheless, after the field application, the induced changes are not lasting permanently. In Fig. 3, there is a texture before and after the electric field application. It is remarkable that after the electric field application, the borderlines between new domains are not so strictly perpendicular to the rubbing direction. It means that the disclination lines are now straightened in this direction. Additionally, the point-like defects are disappearing under the applied electric field.

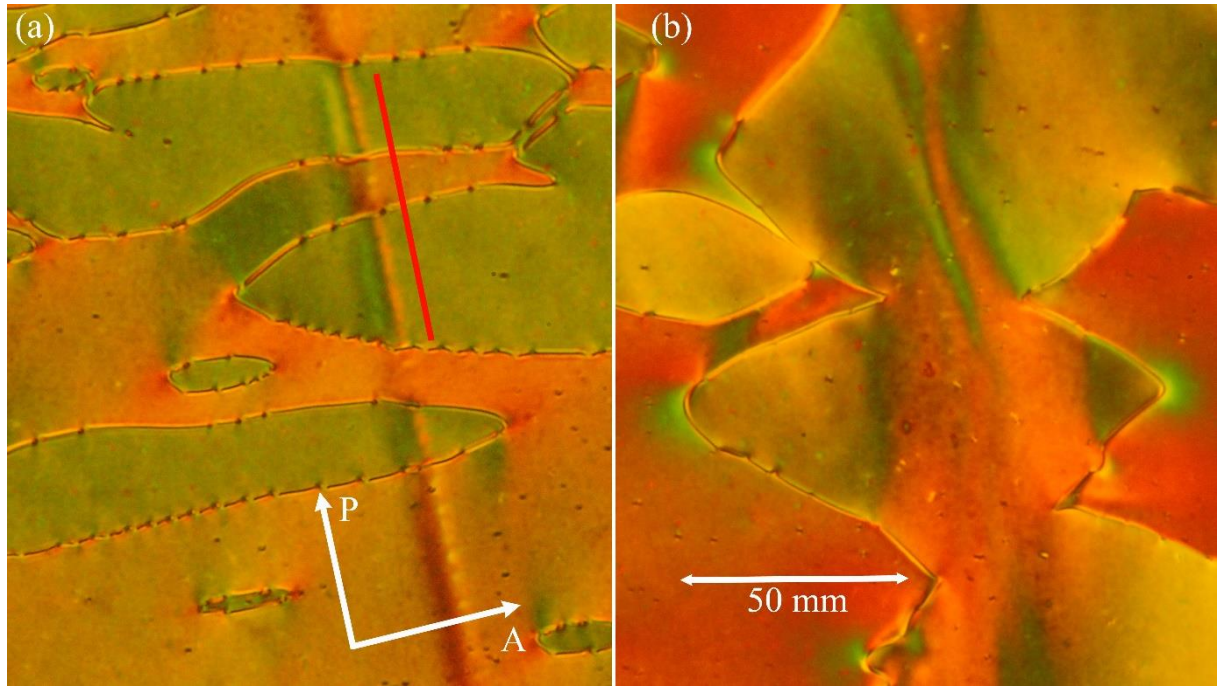

**Fig. 3.** The textures before (a) and after (b) an external field application. The edge of the electrodes lies along the red line, which is marking the rubbing direction. The polariser and analyser are in a perpendicular position, slightly rotated with respect to the photo orientation with the aim to better distinguish the twisted domains.

## 3 Theoretical descriptions

**2π-twist disclination in a finite sample**

In the next part, we will treat 2π-twist disclination as a boundary between the domains of the opposite twist in an elastic approach. The influence of a possible charge will be commented. In Fig. 4, we prepared two schematic pictures to illustrate the twisted domains in two orientations. In Fig. 4a, there is a picture we observe predominantly during our observations, in which the domain borders are perpendicular to the rubbing direction (molecular orientation on the sample surface). In Fig. 4b we show an orientation of the disclination along the rubbing direction for illustration. In the following text, we will try to describe the defect structure theoretically and analyse the obtained models with respect to the observed experimental results. Let us apply the coordinate axes $x$ and $y$ parallel to surface of the glass plates, while the $z$-axis is perpendicular to surfaces (Fig. 4).

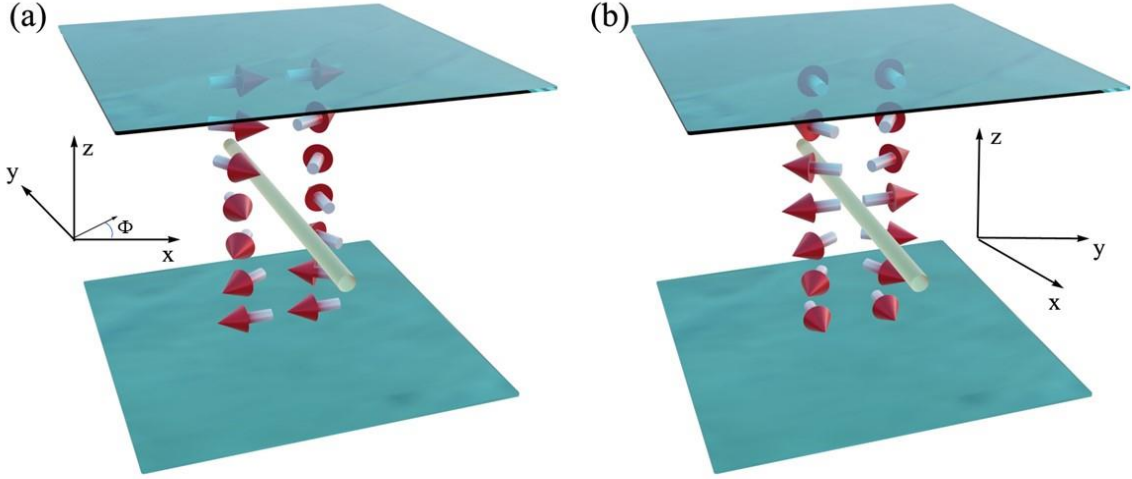

**Fig. 4.** Coordinate axes *x* and *y* lie in the plane of the sample glass plates situated at $z = \pm\frac{h}{2}$, *z*-axis is perpendicular to the glass plates. Director $\vec{n}$ lays in $(x, y)$-plane and is inclined at an angle $\Phi$ from *x*-axis. (a) A disclination line is oriented along *y*-axis and is perpendicular to the orientation of directors at surfaces. The disclination configuration charge is zero. (b) A disclination line is oriented along *x*-axis and is parallel to the director at the surfaces. The disclination configuration charge is non-zero.

For simplicity of the treatment, the director $\vec{n}$ lays in the plane of the sample surfaces and it has components $\vec{n} = (cos\Phi, sin\Phi, 0)$. Angle $\Phi$ is an angle between $\vec{n}$ and *x*-axis in $(x, y)$-plane. The elastic energy density of a nematic liquid crystal can be written in the form [24-26]:

$$\varrho f = \frac{K_1}{2}(div\ \vec{n})^2 + \frac{K_2}{2}(\vec{n}.rot\ \vec{n})^2 + \frac{K_3}{2}(\vec{n} \times rot\ \vec{n})^2\ . \qquad (1)$$

The parameters $K_1$, $K_2$ and $K_3$ are elastic constants for splay, twist and bend, respectively. Moreover, we can expect $K_1 \approx K_3 \approx K$ as it was assumed in literature for the $N_F$ phase [13]. In the $N_F$ phase, the macroscopic polarisation $\vec{P}$ is parallel to $\pm\vec{n}$. Let $\vec{P} = P_o\vec{n}$ with $P_o$ the value of the mean macroscopic spontaneous polarisation. Then the defects such as disclinations or point defects could create a total charge $Q$ [27]:

$$Q = \iiint \rho dV, \qquad (2)$$

where $\rho = -div\ \vec{P}$ is the charge density created by the polarisation distribution around the defect. In formula (2) we integrate over the volume containing the defect. The charge $Q$ can be called "configuration" charge as it depends on the distribution of polarisation (i.e. director) around the disclination. The charge of the defects leads to the energy increase.

As the whole sample has no charge, the local charges of the defects should be compensated both by oppositely charged defects and by free charges in the sample. The discussion in literature [15] indicates a possible electric response of the structure. Nevertheless, the compensation due to the presence of oppositely charged defects is not excluded. A special example is a closed $2\pi$-twist disclination loop (Fig.2). The opposite segments of the loop have opposite configuration charges. Globally, the charge of the loop is zero because the charged segments of the closed loop mutually compensate.

To simplify the elastic energy density (1), let us assume that $K_1 \approx K_3 \approx K$. With a planar director described by angle $\Phi$, the elastic energy density gives the equilibrium equation in the form:

$$\frac{\partial^2 \Phi}{\partial x^2} + \frac{\partial^2 \Phi}{\partial y^2} + \alpha^2 \frac{\partial^2 \Phi}{\partial z^2} = 0, \qquad (3)$$

with $\alpha = \sqrt{\frac{K_2}{K}}$. The glass plates covering the sample are placed at $z = \pm \frac{h}{2}$ with $h$ being the sample thickness. Let $2\pi$-twist disclination be oriented along $y$-axis and situated at the position $z = d$, where $d \in \left(-\frac{h}{2}, \frac{h}{2}\right)$. Then $\Phi$ is the function of the coordinates $x$ and $z$ only and the solution of e.g. ($-2\pi$)-twist disclination can be determined using the method from [28,29] in the form:

$$\Phi = C + arctan\left(tanh\left(\frac{\pi x}{2h\alpha}\right) cot\left(\pi\left(\frac{z}{2h} - \frac{d}{2h}\right)\right)\right)$$
$$+ arctan\left(tanh\left(\frac{\pi x}{2h\alpha}\right) cot\left(\pi\left(\frac{z}{2h} - \frac{1}{2} + \frac{d}{2h}\right)\right)\right). \qquad (4)$$

The constant $C = 0$ for $z > d$ and $C = \pi$ for $z < d$; we can write $C = \frac{\pi}{2}(1 - sgn(z - d))$. This choice of the contant $C$ permits the satisfaction of the boundary conditions $\Phi = 0$ at $z = \frac{h}{2}$ and $\Phi = \pi$ at $z = -\frac{h}{2}$. Note also that for $z = d$ there is a discontinuity in $\Phi$, which is $-2\pi$ for $x < 0$, i.e. $\lim_{z \to d+} \Phi - \lim_{z \to d-} \Phi = -\pi + \pi\, sgn(x) = -2\pi$. For $x > 0$ there is a continuity in $\Phi$. The topological charge of the disclination is defined as an integral taken over the curve enveloping the disclination line in $(x, z)$- plane and centered at $x=0$ and $z=d$. Based on (4), the integral $\oint d\Phi$ can be written as:

$$\oint d\Phi = \left[\lim_{z \to d-} \Phi\right]_{x>0} - \left[\lim_{z \to d-} \Phi\right]_{x<0} + \left[\lim_{z \to d+} \Phi\right]_{x<0} - \left[\lim_{z \to d+} \Phi\right]_{x>0} = -2\pi.$$

The elastic self-energy $E_{el}$ per unit length of the disclination along $y$-axis can be evaluated using the analogous procedures to a previously published method [29]. Since the upper and lower glass plates are equivalent, i.e. they have the same anchoring energy, the disclination is repulsed equally from both surfaces and the disclination is situated in the middle of the sample. Then $d = 0$ and $E_{el}$ can be expressed as:

$$E_{el} = -\pi\sqrt{K_2 K}\ln\left(\sin\frac{\pi r_o}{h}\right), \quad (5)$$

where $r_o$ is a core radius of the disclination. Energy $E_{el}$ is positive because always $h \gg r_o$. It means that disclination loops tend to shorten their circumferences in order to decrease their total self-energy.

Now, the total charge of $2\pi$-twist disclination will be tested. The formula (2) can be transformed into the integral over the surface enveloping the disclination [30,31]:

$$Q = -\iint_S \vec{P}d\vec{S} = -\oint \vec{P}\vec{N}l_y ds, \quad (6)$$

where $l_y$ is the characteristic length of the disclination and $\vec{N}$ is the normal to the surface around the disclination. In expression (6) we integrate over the curve in $(x,z)$-plane, as all functions do not depend on $y$-coordinate. We choose the integration curve enveloping the disclination in the form of rectangle centred to the disclination line. The normal $\vec{N}$ is then parallel to $x$-axis with components $(\pm 1,0,0)$ or parallel to $z$-axis with components $(0,0,\pm 1)$. For $\vec{n} = (\cos\Phi, \sin\Phi, 0)$ the integral (6) from $-R/2$ to $+R/2$ transforms into the form:

$$Q = -\int_{-R/2}^{R/2} P_o l_y \cos\Phi(x\to\infty, z)dz + \int_{-R/2}^{R/2} P_o l_y \cos\Phi(x\to -\infty, z)dz, \quad (7)$$

where $R < h$. The sign of the charge depends on the sense of the polarisation rotation around the closed circuit encircling the disclination. For $x\to\infty$, we obtain: $\lim_{x\to\infty}\Phi \approx \frac{\pi}{2} - \frac{\pi z}{h}$, while for $x\to -\infty$: $\lim_{x\to -\infty}\Phi \approx \frac{\pi}{2} - \pi\,sgn(z) + \frac{\pi z}{h}$. When the disclination line is oriented along $y$-axis, the disclination charge is zero. However, when the disclination line is inclined from $y$-axis at an angle $\varphi$, $\Phi$ is changed to $(\Phi - \varphi)$ and the integration (7) yields:

$$Q = -P_o l_y \frac{4h}{\pi}\sin\left(\frac{\pi R}{2h}\right)\sin(\varphi). \quad (8)$$

A charged disclination segment should be compensated by an oppositely charged segment in the sample bulk. On the other hand, the compensation of small charge perturbations can be also done by induction of the splay deformation, as was recently reported[15] for another compound.

Expression (8) shows that for the disclination line perpendicular to the anchoring direction, i.e. for $\varphi = 0$, the disclination charge is zero (Fig. 4a). For a disclination line inclined from $y$-axis, there is a non-zero disclination charge proportional to $sin(\varphi)$. In the special case with $\varphi = \frac{\pi}{2}$, the disclination line lies along $x$ axis (Fig. 4b). Therefore, a $2\pi$-twist disclination loop is elongated in the direction perpendicular to the rubbing direction to minimise the energy with respect to an electric charge. Nevertheless, the parts of the loop near its tips are charged and have propensity to attract themselves. The contraction of the loop is also favoured by the positive self-energy (5), acting as a line tension of the disclination and forcing the loop to decrease its length. Such a behaviour of disclination loops bounding the twist domains of the opposite twist can be seen in Fig. 1. The elongated segments of the disclinations are mostly oriented in the direction perpendicular to the rubbing direction, which helps to reduce the disclination configuration charge. When the disclination segments are oriented in unfavourable directions, they tend to decrease their length, or they can annihilate (Fig. 2).

**Point-like defects in the core of 2π-twist disclination**

The observations of the twisted domains in the $N_F$ phase show point-like defects decorating disclination lines. Such point-like defects were created simultaneously with the disclination appearance on cooling from the isotropic phase as it was seen from the observation of small twist domains, which diminished and decreased their surface. The existing point-like defects annihilated with vanishing disclination line encircling the domain (Fig. 1). Therefore, we expect the point-like defects relate to the director rotations in the disclination core. Such a director rotation in the disclination core can be in principally non-planar. We will not construct an exact core structure of 2π-twist disclination as it would be rather complicated. For the purposes of our discussion, we suppose that the projection of the director rotations in the disclination core is into $(x, y)$ −plane, i.e. the rotation of the director projection will occur mainly around $z$ −axis. Then the point-like defects could have the director structure similar to the one shown in Fig. 5, in a similar manner as it was described previously for the nematic phase [32].

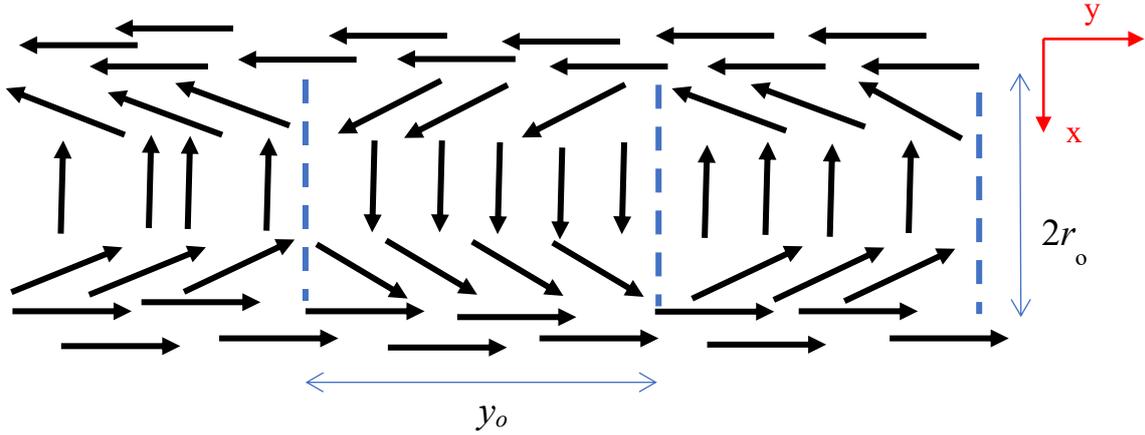

**Fig. 5:** Scheme of the disclination core with the diameter $2r_o$. Arrows are the projections of polarisation onto the plane $(x, y)$. Dashed lines represent projections of infinitesimal disclination loops within the core lying in the plane $(x, z)$, with $z$- axis perpendicular to the plane $(x, y)$. The disclination loops are situated at a distance $y_0$ along $y$-axis.

Due to the topological reasons, the closed disclination loop contains the even number of point-like defects. This can be also deduced from Fig. 5, see an analogy with the classical nematic phase [32]. The core dimensions are characterised by the core radius $r_o$, which can be estimated to be about 0.5 μm according to the resolution of the textural observations. It is smaller than the sample thickness $h$ (of the order of 5 μm). Then we will describe the point-like defect in the disclination core using the concept of infinitesimal disclination loops. This concept was introduced in the dislocation theory [33] for the description of small point-like impurities. This concept was applied previously [34,35] for the description of point-like defects in liquid crystals characterised by the director rotation. In this contribution, the concept of the infinitesimal disclination loops is also applied for an approximative description of a core structure.

The interaction energy $E_I$ of two neighbour point-like defects in the disclination core [36] can be described by the expression:

$$E_I = \Omega_z \Omega_z' K K_2 \oint_C \oint_{C'} G_{zz}(\vec{r} - \vec{r'})(dxdx' + dydy' + dzdz'/\alpha). \qquad (9)$$

In expression (9) $\Omega_z$ and $\Omega_z'$ are $z$−components of Frank vectors characterising the infinitesimal disclination loops $C$ and $C'$, respectively. Parameter $G_{zz}(\vec{r} - \vec{r'})$ is $z$−component of Green function giving the rotation of the director at the point $\vec{r}$ due to $z$−component of a point-force unit moment at $\vec{r'}$. The details of the evaluation of $G_{zz}(\vec{r} - \vec{r'})$ can be found in [34,36] and

$$G_{zz}(\vec{r} - \vec{r'}) = \frac{1}{4\pi K \alpha}\left[(x - x')^2 + (y - y')^2 + \left(\frac{z-z'}{\alpha}\right)^2\right]^{-\frac{1}{2}}. \qquad (10)$$

Interaction energy (9) can be evaluated simply when the infinitesimal loops have radii of the disclination core radius $r_o$. In this case the integrals over $C$ and $C'$ can be approximated by the product of the integrated function taken at the centre of the infinitesimal loop and its length $(2\pi r_o)^2$, which is often called the mean value of the integral. When we consider $2\pi$-twist disclination lying along $y-$ axis, let us centre the infinitesimal loops at points $(0,0,0)$ and $(0, y_0, 0)$ where the distance $y_0$ is positive. Then $G_{zz}(\vec{r} - \vec{r'}) = \frac{1}{4\pi\sqrt{KK_2}}\frac{1}{|y_0|}$ and $E_I$ can be approximated as:

$$E_I = \pi \Omega_z \Omega_z' \sqrt{KK_2} \frac{r_o^2}{|y_0|}. \qquad (11)$$

Components $\Omega_z$ and $\Omega_z'$ should lie in the interval $(0, \pi)$. As neighbouring point-like defects have opposite signs, we write $\Omega_z' = -|\Omega_z|$. Then the elastic interaction force $f_y$ acting between neighbouring point-like defects can be evaluated as:

$$f_y = -\frac{\partial}{\partial t}\left(-\pi \Omega_z^2 \sqrt{KK_2}\frac{r_o^2}{|y_0|}\right) = -\pi \Omega_z^2 \sqrt{KK_2}\frac{r_o^2\, sgn(y_0)}{y_0^2} = -\pi \Omega_z^2 \sqrt{KK_2}\frac{r_o^2}{y_0^2}. \qquad (12)$$

because $y_0 > 0$.

We consider that the neighbouring point-like defects in a disclination core form an unstable configuration and they can attract each other. This analysis was done for an approximative elastic description of the core structure [33]. As we do not know the exact director distribution in the disclination core, we cannot evaluate the possible electric contribution to this interaction. If such a contribution exists, we expect the point-like defects will have also opposite electric charges which affects the elastic attraction. Thus, such configuration is not stable and any external stimuli, e.g. an applied electric field, can remove these point-like defects.

It is necessary to point out that the core structure with the director rotation in plane has no charge. However, when a disclination is charged, one can expect a rotation of non-planar direction in the core to compensate a part of the disclination charge (if not all charge). Similar discussion was presented previously for another compound with the $N_F$ phase [15]. Additionally, in our case of the $N_F$ phase we observed the point-like defects, which represent the director perturbations in the disclination core. It is seen in Fig. 2, where annihilation of the disclination segments with point-like defects is experimentally demonstrated. One can see how the point-like defects cease to exist together with the disclination segments.

## 4 Conclusions and discussion

In this contribution, we present the experimental observation of the textures in the ferroelectric nematic phase, which exists on direct cooling from the isotropic liquid. The observed twisted domains which are created during the phase transition from the isotropic to the ferroelectric nematic phase were analysed and theoretically described. We discussed the character of the disclination charge and concluded that the defects are $2\pi$-twist disclination lines, and they are preferentially orientated perpendicular to the rubbing direction to eliminate the energy cost.

Additionally, we observed the point-like defects decorating these disclination lines, erasing simultaneously during the phase transition from the isotropic phase. We have analysed and described this kind of defects in a disclination core. We can conclude that the mutual interaction of the point-like defects on the observed disclination line is attractive. Therefore, the point-like defects are unstable and can annihilate with a speed depending on the viscosity of the material. Application of an external electric field in the direction perpendicular to the glass surfaces supports the annihilation of these point-like defects. This can be explained by the energy stimulus, which helps to reconnect the neighbouring twisted domains and to reduce the disclination line length. Then disclination lines are oriented in the directions with higher energy and such line configurations are also unstable. Unfortunately, the effect of the applied electric field in this geometry is rather complex and it is difficult to analyse it. For a better analysis, more sophisticated experimental procedures and new cell constructions are necessary to allow for the domain observation under the electric field applied in $(x, y)$-plane.

## Acknowledgements

Authors acknowledge the financial support of the Czech Science Foundation, project 22-16499S.## References:

1. O.D. Lavrentovich. PNAS, 2020, 117 (26) 14629-14631.
2. H. Nishikawa, K. Shiroshita, H. Higuchi, Y. Okumura, Y. Haseba, S.-I. Yamamoto, K. Sago and H. Kikuchi, Adv. Mater., 2017, 29, 1702354.
3. R. J. Mandle, S. J. Cowling and J. W. Goodby, Phys. Chem. Chem. Phys., 2017, 19, 11429–11435.
4. R. J. Mandle, S. J. Cowling and J. W. Goodby, Chem. – Eur. J., 2017, 23, 14554–14562.11